Updated version 2026.09.24 (accepted version)

# Tritium for Nanoscale Hydrogen Analysis by Atom Probe Tomography

Maria Vrellou[1*], Alexander Welle[2,3], Stefan Wagner[1], Marco Weber[1], Rolf Rolli[1], Hans-Christian Schneider[1], Astrid Pundt[1,2], Xufei Fang[1*], Christoph Kirchlechner[1]

[1]Institute for Applied Materials, Karlsruhe Institute of Technology, Kaiserstr. 12, 76131 Karlsruhe, Germany

[2]Karlsruhe Nano Micro Facility (KNMFi), Karlsruhe Institute of Technology, Kaiserstr. 12, Germany

[3]Institute of Functional Interfaces, Karlsruhe Institute of Technology, Kaiserstr. 12, 76131 Karlsruhe, Germany

***Corresponding authors:** maria.vrellou@kit.edu (M. V.); xufei.fang@kit.edu (X. F.)

## Abstract

Accurate nanoscale detection of hydrogen is essential for understanding hydrogen-related phenomena in materials, yet conventional tracing with deuterium is often complicated by residual background hydrogen. This study evaluates tritium as a highly resolvable isotopic marker for nanoscale hydrogen analysis in metals using atom probe tomography. Titanium was selected for its ability to incorporate hydrogen isotopes, providing a suitable platform for tritium detection. Time of flight secondary ion mass spectrometry and electron backscatter diffraction were performed prior to tritium charging to characterize the initial composition and microstructure. Atom probe tomography in laser mode before and after tritium charging, at three post-charging time intervals, enables tracking of tritium incorporation over time. Thermal desorption analysis confirmed the presence of tritium and complemented the secondary ion mass spectrometry measurements, highlighting the role of the surface oxide layer in modulating tritium release. While tritium, deuterium, and protium differ in their diffusion and trapping behavior, the distinct mass signal associated with tritium provides a practical advantage for resolving hydrogen at very low concentrations. This work serves as a fundamental benchmarking study for leveraging tritium and atom probe tomography as a combined tool for understanding hydrogen in materials, relevant for interpreting local processes such as hydrogen embrittlement.

**Keywords:** Atom probe tomography, hydrogen isotopes, tritium, titanium, thermal desorption analysis

## 1. Introduction

Hydrogen and its isotopes protium, deuterium, and tritium are central to the energy transition. They hold promise as carbon-free fuels for future energy systems and as essential elements in fusion reactors (Luís et al., 2021; Villari et al., 2025). At the same time, their interaction with structural metallic materials such as high-strength steels can severely compromise the mechanical integrity, a phenomenon known as hydrogen embrittlement (HE) (Borchers et al., 2008; Dantzer, 1997; Hong et al., 2022; Lynch, 2012; Murakami et al., 2010; Robertson et al., 2015). This degradation process threatens the safety and durability of critical infrastructures, making it vital to understand hydrogen behavior in advanced microstructures, whose properties critically depend on defect structures such as interfaces, dislocations, and point defects (Chen et al., 2017; Kirchheim, 2010; Lynch, 2012; Murakami et al., 2010; Pundt and Kirchheim, 2006; Robertson et al., 2015; Takahashi et al., 2012).

Atom probe tomography (APT) enables near-atomic-scale mapping of hydrogen and its isotopes, and remains unique in providing three-dimensional spatial and compositional information with quasi-atomic spatial resolution within materials (Chen et al., 2017; Gault et al., 2024, 2021; Takahashi et al., 2012). Nevertheless, accurately quantifying hydrogen using APT remains a major challenge due to the combined effects of hydrogen's high mobility, its ubiquitous presence in experimental environments (Mayweg et al., 2024, 2023; Sundell et al., 2013), and several intrinsic limitations of the APT technique (Chen et al., 2023; Kolli, 2017a).

The APT sample preparation procedures may alter the hydrogen content of the specimen. Hydrogen's interstitial nature and exceptionally high diffusivity in metals even at low temperatures make it prone to rapid ad/absorption and desorption, especially at ambient and higher temperatures. Consequently, specimen preparation (mechanical polishing, electropolishing, or focused ion beam (FIB) milling) and transfer steps can alter the hydrogen content and distribution (Mayweg et al., 2024).

Namely, electropolishing has been reported to introduce additional hydrogen into APT specimens (Breen et al., 2020; Y. Chang et al., 2019; Wang et al., 2024), whereas room-temperature FIB milling is known to cause hydrogen uptake and even hydride formation in sensitive alloys such as Ti and Zr (Y. Chang et al., 2019; Mayweg et al., 2023, 2024; Yan et al., 2019). Hydrogen-charged specimens may also undergo rapid hydrogen desorption during post-charging handling (Devaraj et al., 2023). These issues are exacerbated in the small, needle-shaped APT specimens, where the high surface-to-volume ratio accelerates hydrogen loss (Suzuki and Takai, 2012). To mitigate these effects, cryogenic FIB preparation has been adopted to suppress hydrogen uptake and prevent hydride formation during specimen milling (Breen et al., 2020; Chang et al., 2018; Y. Chang et al., 2019; Koç et al., 2025; Mouton et al., 2021). Vacuum-cryogenic transfer systems using dedicated shuttles are employed to minimize outgassing and the ingress of atmospheric moisture during specimen handling (Chen et al., 2017; McCarroll et al., 2020).

Vacuum-cryogenic transfer systems using dedicated shuttles are also employed to reduce hydrogen redistribution during specimen handling. Devaraj et al. and Vukkum et al. (Devaraj et al., 2023; Vukkum et al., 2025) demonstrated that transferring plunge-frozen, hydrogen-charged APT needles under vacuum conditions preserves their hydrogen content by preventing rapid desorption at ambient temperature. However, the extent to which hydrogen retention arises from cryogenic temperatures, vacuum conditions, or their combination remains material-dependent. Surface oxides can either inhibit or promote hydrogen loss depending on their reducibility. For example, in Mg hydrogen remains stable under vacuum but is depleted when oxygen is present due to water-forming reactions at the surface (Uchida et al., 2015).

Even when specimens are prepared carefully, additional complications arise from the analysis environment itself. Despite the ultra-high vacuum (UHV) conditions in the APT analysis chamber, a certain level of residual gas remains, that needs to be considered when analyzing APT data. Among these species, hydrogen is typically the dominant component, mainly originating from the continuous outgassing of the stainless-steel chamber walls and in-vacuum parts of the atom probe instrument (Felfer et al., 2022; Gupta et al., 2015; Hong et al., 2008; Mayweg et al., 2024; Meier et al., 2021). The residual $H_2$ molecules may adsorb onto the surface of the APT specimen, where they are ionized and evaporated under the strong electrostatic field at the tip apex (Mayweg et al., 2023; Sundell et al., 2013). This process produces $H^+$, $H_2^+$ (Chen et al., 2020, 2023; Gault et al., 2024; Kolli, 2017b), and sometimes (depending on several parameters such as the experimental protocol and the material) even $H_3^+$ ions (Chen et al., 2020; Ernst and Block, 1983.; Reckzügel et al., 1995; Stepień and Tsong, 1998), which appear in the mass-to-charge-state ratio spectra as peaks at 1, 2, and 3 Da, respectively.

The 1 Da peak corresponds to protons ($H^+$), which may arise via (i) direct ionization of a H atom, (ii) double ionization of the molecule followed by field dissociation into two $H^+$ ions, or (iii) single ionization leading to the formation of $H^+$ and a neutral $H^0$ atom, the latter being undetectable in APT. As a result, the 1 Da signal is predominantly attributed to residual background hydrogen (Chen et al., 2017; Felfer et al., 2022; Mayweg et al., 2024, 2023; Sundell et al., 2013; Takahashi et al., 2012). The 2 Da peak corresponds to molecular $H_2^+$ ions produced by the direct ionization of adsorbed $H_2$. A much smaller contribution to the 2 Da peak may arise from $D^+$ (Takahashi et al., 2010), whose natural isotopic abundance is only 0.0145 at.% (Gault et al., 2024). The formation of $H_2^+$ is particularly favored under laser-pulsing conditions, as the lower effective electrostatic fields associated with laser assistance reduce the likelihood of molecular dissociation into protons (Breen et al., 2020; Chen et al., 2020, 2019; Kim and Seidman, 2015; Mayweg et al., 2024, 2023; Tsong et al., 1983). The 3 Da peak corresponds to $H_3^+$. While neutral $H_3$ is unstable in free space, $H_3^+$ is a stable molecular ion and a field-induced, surface-mediated species involving interactions between adsorbed H and $H_2$ under high electrostatic fields according to the reaction sequence (Ernst and Block, 1983.; Reckzügel et al., 1995; Stepień and Tsong, 1998; Tsong et al., 1983):

$$H_2 + H \rightarrow H_3 \rightarrow H_3^+ + e^- \quad (1)$$

Under the APT electrostatic field, the transient $H_3$ complex can be stabilized and ionized upon formation, allowing $H_3^+$ to be desorbed and detected before field-induced dissociation occurs (Kolli, 2017b). The observation of $H_3^+$ ions in the mass spectra is sensitive to experimental parameters such as field strength, laser pulse energy, and specimen apex temperature (Kolli, 2017b). High-voltage pulsing can be used to suppress $H_2^+$ and $H_3^+$ formation and amplifies the $H^+$ signal at 1 Da by dissociating molecular hydrogen into $H^+$ (Y. H. Chang et al., 2019; Takahashi et al., 2018). Charge-state ratio monitoring of the majority lattice element serves as an in situ indicator of local electrostatic field strength and aids in the interpretation of hydrogen-related peaks (Breen et al., 2020; Diagne et al., 2024; Kingham, 1982; Shariq et al., 2009; Tegg et al., 2024). Moreover, the use of titanium-alloy chamber components and getter materials has been shown to minimize residual hydrogen within the analysis environment (Felfer et al., 2022). In addition to these residual gas contributions, hydrogen originally present in solid solution within the specimen may also contribute to the detected hydrogen peaks, as demonstrated by Meier et al. (Meier et al., 2021). They showed that lattice-bound hydrogen isotopes (studied via deuterium as a tracer) produce atomic ions that persist even when contaminant hydrogen is minimized, indicating that a fraction of the corresponding hydrogen isotope signal may arise from hydrogen dissolved in the material.

Apart from the well-known diffusion mechanisms, athermal tunneling-mediated migration of hydrogen (Gault et al., 2024) can also influence the measured distributions. Quantum tunnelling allows light atoms to penetrate energy barriers even when thermal activation is suppressed, and tunnelling effects become increasingly significant at low temperatures, where classical over-the-barrier hopping slows down while tunnelling rates remain substantial (Kästner, 2014). The tunnelling probability decreases exponentially with increasing particle mass, as described by the Wentzel–Kramers–Brillouin (WKB) approximation (Griffiths, 1995), so protium exhibits the highest tunnelling propensity, whereas deuterium and tritium show severely reduced tunnelling probabilities. In BCC metals, the short jump distances between tetrahedral sites make tunnelling most favorable, and protium tunnelling becomes relevant just below 0 °C and is expected to persist down to cryogenic temperatures due to zero-point motion (Fukai, 2005). In FCC metals, larger octahedral-site barriers further reduce tunnelling and in HCP metals such as Ti, the tightly packed interstitial geometry and higher migration barriers make tunnelling even less favorable. As a result, isotope-dependent mobility and trap occupancy may be modified during specimen preparation and cold-stage handling, and isotope effects arising from tunneling should therefore be considered when interpreting APT data. So, although cryogenic preparation, transfer, and APT analysis conditions strongly suppress thermally activated diffusion, quantum-driven migration cannot be entirely excluded.

To overcome the limitations imposed by residual hydrogen, isotopic labelling is often employed. Deuterium ($^2H$) is widely used as a practical surrogate for hydrogen (Gemma et al., 2007; Takahashi et al., 2010) because of its lower diffusion coefficient and its significantly lower tunnelling probability (Maxelon et al., 2001), making deuterium less prone to migrate within the nano-specimen especially during the analysis at low temperatures (Gemma et al., 2009). Indeed, previous APT studies have demonstrated that deuterium distributions consistent with expected trapping behavior can be measured at temperatures below ~30 K (Gemma et al., 2012). However, the detection of deuterium ($^2H^+$) is often complicated by the limited mass resolution of APT, which prevents a clear distinction between deuterium peaks and the background signal from the molecular ion $H_2^+$ mainly originating from residual hydrogen. In contrast, such a distinction can be achieved using ToF-SIMS, owing to its considerably higher mass resolution compared to that of APT, although this comes at the expense of quantitative accuracy. Common SIMS setups, independent of the mass filter being a time-of-flight or a magnetic sector field analyzer, can provide a mass resolving power up to 5000 or even 10000 M/ΔM, as discussed extensively in (Li et al., 2020). In the used instrument a high mass resolution is achieved by “bunching” the primary ion pulses. The liquid metal ion gun uses a precisely timed axial electric field to provide an offset on the kinetic energy of the primary ions (25 keV) such that the pulse width of the beam decreases from >10 ns down to <1 ns before the primary ions are impinging the sample surface. This allows for a rather high target current, but the introduced energy spread results in polychromatic aberration, reducing the lateral resolution during imaging which was not of concern for our present study.

To address these analytical challenges, tritium ($^3H$) has recently been employed as an isotopic tracer in APT studies, including the investigation of tritium content and neutron irradiation effects in $LiAlO_2$ pellets used in tritium-producing burnable absorber rods (TPBARs) (Devaraj et al., 2021), as well as in investigations of HE in twinning-induced plasticity (TWIP) steels (Khanchandani et al., 2023). Its higher atomic mass, low diffusivity, and extremely low natural abundance (about $10^{-18}$ at.% ("Tritium | Radioactive, Hydrogen, Decay | Britannica," 2025)) render its presence from residual gases in the analysis chamber negligible. At the same time, tritium handling requires licensed radiological facilities and dedicated safety infrastructure, making its use more complex than deuterium. The present work aims to validate tritium as a highly resolvable isotopic marker for directly studying hydrogen behavior in metals at the nanoscale using APT, representing an essential step toward enabling reliable quantification of hydrogen isotopes in future APT investigations. We note that tritium, deuterium, and protium differ in diffusivity, trapping strength, and phase stability, and these isotopic differences cannot be fully resolved within the scope of this study. Thermal desorption analysis (TDA) was additionally performed to independently verify the global presence of tritium in the specimens by monitoring its release during controlled heating. Commercially pure titanium was selected as a model material because it readily absorbs hydrogen isotopes and provides conditions favorable for detecting even low levels of tritium. While titanium readily incorporates hydrogen isotopes, gas charging is not equally effective for alloys with intrinsically low hydrogen solubility (Chen et al., 2017). In such materials, achieving appreciable hydrogen uptake requires a large chemical potential, which can be provided either by high pressures and elevated temperatures during gas charging or by electrochemical charging at room temperature using sufficiently high cathodic potentials and recombination poisons that hinder hydrogen recombination and desorption before absorption. Electrochemical charging, however, introduces additional challenges. Corrosive electrolytes often necessitate protective coatings, such as Pd, which also promote hydrogen absorption (Guardi et al., 2024). In addition, recombination poisons are highly toxic (Tiegel et al., 2016). Consequently, the experimental conditions for electrochemical charging can be more challenging than those for gas charging. These considerations indicate that extending tritium-based nanoscale hydrogen analysis to materials with low hydrogen solubility may require alternative charging strategies. The advantages and possible limitations of this tritium-based approach are discussed later.

## 2. Materials and method

### 2.1. Sample preparation and EBSD analysis

A commercially available titanium rod (99.9+% purity, CAS No. 7440-32-6) 100 mm long and with a diameter of 6.35 mm from MaTeck (Material, Technologie & Kristalle GmbH, Jülich, Germany) was used in this study. Disk-shaped samples, each ~0.7 mm in thickness, were cut from the rod. These disk-shaped specimens were mechanically polished using a QATM Saphir 520/Rubin 500 system. Grinding was carried out sequentially with silicon carbide (SiC) abrasive papers of progressively finer grit sizes (from #320 to #4000), followed by polishing with a suspension containing 9 µm diamond particles. Final surface finishing was achieved using colloidal silica (OP-S) mixed with a few drops of hydrogen peroxide ($H_2O_2$). Grain size was determined by electron backscatter diffraction (EBSD) using an Oxford Instruments Symmetry S2 detector, at acceleration voltage of 20 kV and a current of 4 nA in a ZEISS Crossbeam 550 L FIB/SEM. EBSD data were acquired using the AZTEC software, and the resulting datasets were subsequently processed and analyzed using AZTEC Crystal as well as a custom MATLAB script based on the MTEX toolbox.

### 2.2. ToF-SIMS

Time-of-flight secondary ion mass spectrometry (ToF-SIMS) was employed prior to tritium exposure, to qualitatively identify the elemental constituents of the material. The analyses were performed under ultra-high vacuum conditions using an IONTOF ToF-SIMS.5 instrument, without targeting a specific grain orientation. A $Bi^+$ primary ion beam operated in high current bunched (HCB) mode, yielding a 1.3 pA current, was used for spectral acquisition to provide high mass resolution and enhanced secondary ion yields. To evaluate the elemental distribution as a function of sputtering time, depth profiles were acquired over 500 µm × 500 µm areas in both positive and negative polarity. At least two independent depth profiles were collected in each polarity.

Positive polarity analyses were performed using a 1 keV $O_2^+$ ion beam scanned over a 800 µm × 800 µm area, 240 nA beam current. For negative polarity measurements 1 keV $Cs^+$ ions scanned over a 800 µm × 800 µm area, 110 nA beam current were used. Signals prone to detector saturation (like $^{16}O^-$) were not considered but replaced by suitable signals of lower intensities (e.g. $^{18}O^-$). Depth profiles are consequently reported as a function of ion fluence rather than nanometers. Additionally, owing to the element and matrix dependence of SIMS sputter yields, the resulting depth scale is material-specific and not necessarily linear. Data acquisition and processing were performed using IONTOF SurfaceLab 7 software.

### 2.3. Charging with a hydrogen isotope gas mixture

Tritium charging[1] (the term charging is used instead of loading to avoid confusion with mechanical loading) was carried out on the bulk disk-shaped specimen, at the Fusion Materials Laboratory (Karlsruhe Institute of Technology, Germany), using a gas mixture of nominally 500 appm tritium ($^3H_2$) in hydrogen ($^1H_2$) (bottled on 24 June 2014, TRITEC Ltd., Switzerland). The decay in tritium concentration over time was estimated using the radioactive decay equation:

$$N(t) = N_0 \times e^{-\lambda t} \quad (2)$$

where $N_0$ is the number of initial tritium nuclei in the gas mixture, $\lambda = \frac{\ln(2)}{t_{1/2}} \approx 1.8 \times 10^{-9}\, s^{-1}$ is the tritium decay constant with $t_{1/2} = 12{,}32\, years = 3{,}89\ \times 10^8 s$ being the half-life of tritium(Canadian Nuclear Safety, 2026.). *t* is the time elapsed between the gas bottling date and the charging experiment (in years), which will be given later.

Three tritium charging campaigns were carried out, followed by TDA and APT analyses. Prior to tritium charging, the charging system was evacuated and purged with helium in two consecutive cycles to remove residual gases before introducing the tritium – hydrogen gas mixture. In all three campaigns, charging was performed at 500 °C and 1 bar for 6 hours under the tritium-containing gas mixture. After charging, samples that were not immediately analyzed, were stored under nitrogen atmosphere (99.2% $N_2$) to minimize surface reactions.

The first charging campaign was conducted in May 2024, when the tritium concentration in the gas mixture had decayed to approximately 281 appm. One titanium sample was charged and analyzed by APT after 150 days of storage for investigating the effect of tritium retention. The second campaign took place in January 2025, when the tritium concentration was estimated to be 260 appm owing to its decay. Three samples were charged. Two were analyzed by APT and TDA 1 day after charging, while the third was analyzed by APT after 7 days. Several of the 1-day APT tips exhibited significant surface contamination (see below) and were therefore excluded from the analysis. To compensate for this loss, a third campaign was carried out in March 2025, when the tritium concentration had further decreased to approximately 254 appm. Samples from this campaign were analyzed by APT after 1 day of charging.

[1] In literature, "hydrogen loading "is often used instead of "hydrogen charging", however, we deliberately use charging here to avoid confusion with mechanical loading carried out by our group in parallel to this work.

### 2.4. Thermal Desorption Analysis

Thermal desorption analysis (TDA) was performed after tritium charging using an ionization chamber (FML, Karlsruhe Institute of Technology, Germany). This TDA setup has been employed previously at KIT for tritium desorption studies (Heuser et al., 2020; Khanchandani et al., 2023; Leys et al., 2024). The detector output was first calibrated with a 2 GBq $^{3}H_2$/He 5.0/$^{1}H_2$ 0.1% calibration gas obtained from TRITEC Ltd., Switzerland in 2018. During calibration, different carrier gas flow rates were used, corresponding to different tritium activities, and the resulting current (of the order of pA) was measured to determine a calibration factor. This factor was then applied to convert the measured current in the carrier gas into tritium activity (Bq).

The background noise was measured just before the desorption measurements started. TDA analysis was conducted under a continuous flow of carrier gas (10 ml/min He with 0.1 vol% $^{1}H_2$). The sample was heated from room temperature to 1100 °C at a rate of 7 °C/min, followed by a 3 h hold at the maximum temperature and subsequently cooled down back to ambient temperature. Tritium atoms in the flow gas travel through an approximately 11 m long pipeline that is heated at 200 °C before reaching the ionization chamber, where they get detected with a detection efficiency of approximately 30%. For each sample, the background activity was subtracted to obtain a background-corrected release rate (Bq/s), which was then used to calculate the tritium release rate and the corresponding global tritium concentration.

However, TDA measurements are highly sensitive and can be influenced by external factors such as nearby equipment and fluctuations in the hot-lab ventilation system. Further, given the high diffusivity of the hydrogen isotopes contained in the purge gas, a fraction of the transported tritium is expected to migrate into the steel piping of the TDA setup. This process may introduce interferences that contribute to the observed elevated background signal at the end of the TDA measurement, thereby limiting the accuracy of the TDA for absolute quantification. Nevertheless, the shape of the TDA release curve still provides reliable physical insights.

### 2.5. Atom Probe Tomography

For both the as-prepared and tritium-charged states, lift-outs were performed at room temperature in the FEI Scios FIB/SEM located in the Fusion Materials Laboratory (KIT), where radioactive samples can be handled. Lift-outs were prepared using standard methods (Lefebvre-Ulrikson, 2016), and multiple regions of each bulk specimen were sampled and welded onto coupon pre-tips.

Because the APT system is located at the Institute of Nanotechnology (KIT), outside the radioactive facility, the final tip-sharpening steps were carried out on the same day in a ZEISS Auriga SEM/FIB situated adjacent to the APT instrument. During transfer between the two facilities, the welded lift-out chunks were exposed to air for approximately 10 to 15 min before being loaded into the Auriga chamber. The annular milling and the final low-voltage $Ga^+$ ion cleaning step (5 kV, 90 pA) were also performed at room temperature. SEM images of APT tips are illustrated in **Figure S1** in **Supplementary Section S1**. The sharpened tips were introduced into the APT airlock within a couple of minutes. Although the close proximity minimizes the delay between FIB preparation and transfer, brief exposure to air during handling is unavoidable and may lead to limited surface oxidation of titanium and possible hydrogen loss. In most analyzed tips, oxygen-related species were uniformly distributed and did not indicate the presence of a distinct oxide layer (**Figure S4**). One tip measured one day after charging exhibited a slightly non-uniform distribution of O and TiO signals (**Figure S5**), as discussed in the **Supplementary Material S3**.

The APT experiments were performed the day after the samples were placed in the APT airlock (corresponding to approximately 14 hours) using a CAMECA LEAP 4000 X HR. The as-prepared samples were analyzed in voltage-pulsing mode at 30 K, with a 20% pulse fraction and 0.3% detection rate, similar to previous hydrogen-related APT studies (Gemma et al., 2022, 2012; Wilke et al., 2011). A pulse rate of 200 kHz was used. Eight successful measurements were collected, yielding on average one million atoms per run. To improve sample survival and data quality, subsequent APT analyses of as-prepared and tritium charged samples were conducted in laser-pulsing mode at 30 K, using 50 pJ laser energy, a 200 kHz pulse rate and 0.3% detection rate. For each sample and charging condition, three to four successful measurements were obtained in laser-pulsing mode, each comprising approximately 3.8 million atoms. The analysis chamber variations were only small and are briefly discussed in **Supplementary Section S1,** along with the effect of the Ga damage to the total hydrogen content.

The obtained data were reconstructed using the CAMECA AP Suite 6.3.2.135. To account for variations due to the dataset size, ion-weighted means and standard errors were calculated for each material state (“Weighted mean,” 2026.; “Weighted standard deviation,” 2026). The spatial distribution of solutes and hydrogen species within the APT tips was evaluated through a nearest-neighbor analysis. Experimental distributions were compared with a theoretical random distribution generated by shuffling the chemical identity of atoms at each coordinate position within the reconstructed volumes. This statistical comparison was complemented by $\chi^2$ tests.

In the mass-to-charge-state ratio spectrum, of particular significance are the peaks at 17 Da and 25.5 Da, which could correspond either to $(^{50}TiH)^{3+}$ and $(^{50}TiH)^{2+}$, respectively, and / or alternatively to vanadium isotopes ($^{51}V^{3+}$ and $^{51}V^{2+}$). In addition, the 17 Da peak may also contain a contribution from $OH^+$, which commonly appears at this mass-to-charge-state ratio in atom probe spectra due to residual gas or surface-related species.

As ToF-SIMS measurements detected no vanadium in the specimens, a vanadium contribution was excluded. To separate TiH from possible $OH^+$ contributions, the intensity ratio of the isolated $^{46}Ti^{3+}$ and $^{46}Ti^{2+}$ peaks (whose mass-to-charge ratios do not overlap with other species) was used as a reference. Assuming that the charge-state ratio $^{46}Ti^{3+}/^{46}Ti^{2+}$ is comparable to that of $(^{46}TiH)^{3+}/(^{46}TiH)^{2+}$ under identical experimental conditions and considering the similar natural abundances of $^{46}Ti$ (8.25%) and $^{50}Ti$ (5.18%), this relationship was used to estimate the expected contribution of $(^{50}TiH)^{3+}$ relative to $(^{50}TiH)^{2+}$. The remaining signal at 17 Da was attributed to $OH^+$. Deconvolution indicated that the $TiH^{3+}$ contribution was minor compared with $TiH^{2+}$, but it was retained in the analysis to improve the accuracy of the peak decomposition.

**Figure 1** illustrates the 15.5 to 26 Da region of mass-to-charge-state ratio spectra of laser runs before charging and for each post-charging interval. With $^{46}Ti^+$ and $^{50}TiH^+$ defined, the remaining peaks can be deconvoluted accordingly using the natural abundance. The peak at 16 Da also contains $^{16}O^+$, which was assigned as the residual signal after accounting for Ti and $TiH^{n+}$ species.

Another interesting region of the mass-to-charge-state ratio spectrum is the peak at 19 Da. Across all datasets, only two laser-run analyses after charging exhibited this peak. One measured 7 days after loading and the other 150 days after loading, and these are also the largest datasets within their respective post-loading intervals. Since these datasets also contain carbon-related peaks, we have originally attributed peak at 19 Da to carbon molecular ions. However, it is important to note that this could also be $^{16}O^3H$ and thus increasing the reported total tritium count.

Since the voltage-mode runs did not contain any $Ti^+$ ions, the charge-state ratio (CSR) was calculated as:

$$CSR\ (Ti) = \frac{n(Ti^{3+})}{n(Ti^{2+})} \tag{3}$$

where $n(Ti^{3+})$ and $n(Ti^{2+})$ are the detected ion counts of $Ti^{3+}$ and $Ti^{2+}$, respectively. To ensure that the ion counts originated solely from titanium, only the $^{46}Ti$ isotope was used, thereby avoiding overlaps with peaks from other species. The corresponding electric field was then estimated using the formula as proposed by Tegg et al. (Tegg et al., 2024):

$$F = a\left(1 - \frac{b}{CSR^{0.3} + b + 0.256}\right) \quad (4)$$

with $a$ = 16.24 and $b$ = 0.1968 determined for the $Ti^{3+}/Ti^{2+}$ CSR. For each analysis dataset, the CSR and its associated field were calculated. For each material condition, their mean values and standard errors were determined.

Further, to visualize the variation of the field, the x, y, z coordinates and the mass-to-charge-state ratio of each detected ion were derived from the acquired .pos files and were processed using analysis scripts adapted from the Python code originally provided by Tegg et al. (Tegg et al., 2024). The script ranges each ion and bins it into 0.25 nm × 0.25 nm pixels that integrate over the full tip length, generating a two-dimensional map of the field-evaporated ions. The choice of the pixel size is discussed in **Supplementary Section S2**. The ratio of the ion counts within these bins was then used to calculate and plot the CSR, which was subsequently inserted into Equation (3) to derive and visualize the corresponding local electric-field intensity, thus enabling direct correlation between $^{46}Ti$ charge-states, field strength, and spatial distribution.

## 3. Results

### 3.1. Microstructure prior to tritium charging

EBSD was performed on the disk-shaped Ti specimens prior to tritium charging to quantify the grain size, given that hydrogen adsorption has been reported to exhibit grain-size dependence in vapor-deposited titanium films (Tanaka et al., 2009), although such effects are not anticipated to be significant in bulk material. Inverse pole figure (IPF) maps of the Ti samples (Error! Reference source not found.) reveal grains with a random crystallographic distribution, having an average diameter of 110 ± 21 μm, with the scatter bar corresponding to the standard deviation. Several grains displayed twin boundaries, with misorientation angles close to 64° and 85° between neighboring grains, consistent with known twinning systems in HCP Ti (Zahiri et al., 2023). No titanium hydride phases were detected prior to tritium charging.

### 3.2. Elemental constituents of the sample prior to tritium charging based on ToF-SIMS

ToF-SIMS analysis was performed for the identification of the elemental constituents of the material and depth profiling provided qualitative insight into the surface and near-surface chemistry of the material. Signals from $Ca^+$, $K^+$ and $Cl^-$ were detected at the surface, consistent with residual contaminants introduced during mechanical preparation. $Cu^-$ was also detected, exhibiting a maximum at the sample surface, followed by a rapid decrease, a second peak at approximately $2 \times 10^{16}$ ions·cm$^{-2}$, and finally stabilizing in the bulk (**Figure 3a**). $Al^-$ was present in very small amounts throughout the sample.

Carbon ($C^-$) was also present and increased progressively with sputtering until reaching a stable plateau at ~$6 \times 10^{16}$ ions·cm$^{-2}$, indicative of a relatively uniform carbon concentration beyond the surface region. The $Si^-$ signal displayed a sharp peak at ~$1 \times 10^{16}$ ions·cm$^{-2}$, followed by a rapid decrease and subsequent stabilization after ~$3 \times 10^{16}$ ions·cm$^{-2}$, consistent with a thin silicon-rich surface layer. Depth-integrated ion maps revealed that the overall chemical distribution was largely homogeneous, with only carbon, oxygen and $TiO^-$ showing weak grain-dependent contrast (**Figure 3b).**

A native oxide layer was identified through the behavior of oxygen- and titanium-oxide-related ions, which displayed characteristic maxima at low sputtering fluence. The primary oxygen signal ($^{16}O^-$), as commonly observed, was saturated and therefore lay outside the detector's dynamic range. Consequently, the $^{18}O_2^-$ signal was monitored instead, which remained within the measurable range. Both the $^{18}O_2$- and $TiO_3^-$ signals exhibited pronounced maxima at low sputtering fluence (~2 × $10^{15}$ ions·cm$^{-2}$), followed by a rapid decay, indicating that the oxide is confined to the near-surface region. In contrast, the $TiO^-$ signal increased beyond ~2.5 × $10^{15}$ ions·cm$^{-2}$ and formed a broad plateau before reaching a maximum at ~1 × $10^{16}$ ions·cm$^{-2}$. This apparent broad transition does not indicate intermediate oxidation states but rather arises from ion-beam–induced mixing at the crater bottom together with lateral variations in oxide thickness (**Figure 3b**). The $Ti_2O^-$ and $^{49}TiOH^-$ species displayed bell-shaped distributions with maxima at ~1.4 × $10^{16}$ and ~1.9 × $10^{16}$ ions·cm$^{-2}$, respectively.

ToF-SIMS detected hydrogen-related signals originating from environmental contamination. Peaks were observed at 1 Da and 2 Da, whereas no signal was detected at 3 Da (**Figure 4a**), documenting that tritium is not present in the sample before charging. A minor secondary feature near 1 Da was also visible (**Figure 4b**), which is attributed to possible electronic artifacts. Owing to the high mass resolution of ToF-SIMS, the 2 Da region could be resolved into two distinct peaks, both exhibiting intensities approximately three orders of magnitude lower than the 1 Da peak (**Figure 4c**). The first peak, located at ~2.0145 Da, is consistent with $^2H^+$, while a second, slightly higher-mass peak at ~2.0163 Da can be attributed to $H_2^+$. By comparing the integrated intensities of the 1 Da and ~2.0163 Da peaks, the resulting isotopic ratio corresponds to ≈99.99% $^1H$ and ≈0.01% $^2H$, which is consistent with the natural hydrogen isotopic abundance. In addition, TiOH and TiH molecular ions such as $TiH^-$ and $TiH_3^-$ were also detected. Their depth profiles show a near-surface maximum, followed by a continuous decrease with increasing depth, i.e. fluence. Depth-integrated ToF-SIMS images confirm that TiH molecular ion species are present only in very small amounts relative to the total titanium or hydrogen content and suggest uniform distribution (**Figure 2b**).

### 3.3. Elemental constituents of the sample prior to tritium charging based on APT measurements

The composition of the samples prior to tritium charging was further quantified by APT, conducted in both voltage- and laser-pulsing modes, and is reported in **Table 1**. Because these specimens were not intentionally charged with hydrogen isotopes, the peak at 1 Da and hydrogen-related contributions at 2 Da were excluded from the compositional analysis and attributed to residual environmental hydrogen. No peak at 3 Da was observed in the uncharged specimens. Nevertheless, hydrogen-related molecular ions ($H_{MI}$), i.e., titanium hydride ions such as $TiH^{n+}$, were included in the analysis because they form during field evaporation and provide indirect information on hydrogen associated with titanium.

The titanium content was measured as 98.6 ± 0.4 at.% under voltage pulsing and 99.0 ± 0.1 at.% under laser pulsing. Apart from the typical contaminants introduced during mechanical preparation, APT identified the same minor impurities as detected by ToF-SIMS (namely O, C, Si, Al, and Cu), along with gallium implanted during FIB milling. Nitrogen was observed exclusively in the laser-pulsed datasets, most likely arising from laser-assisted surface desorption rather than bulk incorporation. Localized heating and transient perturbations of the electric field during laser pulsing promote the ionization of surface-adsorbed nitrogen, in contrast to the more stable, field-driven evaporation characteristic of voltage pulsing (Kelly et al., 2014; Shaikh et al., 2026).

To quantify hydrogen and compare it with tritium-charged specimens, the composition was recalculated to include all hydrogen peaks. **Table 2** summarizes the ion-weighted average hydrogen content and the individual contributions from the 1 Da, 2 Da, and 3 Da peaks for both voltage- and laser-pulsed datasets. Signals at 1 Da and 2 Da were consistently detected in both acquisition modes, while no peak at 3 Da was observed in either mode (**Figure 5a**).

It is observed that laser pulsing results in nearly twice the total hydrogen concentration compared with voltage pulsing. However, substantial tip-to-tip variability was evident. In voltage mode, hydrogen contents ranged from 0.31 ± 0.01 to 4.73 ± 0.09 at.%, while in laser mode they ranged from 2.25 ± 0.01 to 6.00 ± 0.02 at.%. These differences were largely driven by fluctuations in the 1 Da peak, which varied from 0.23 ± 0.01 to 3.59 ± 0.13 at.% in voltage mode, and from 1.40 ± 0.01 to 5.41 ± 0.02 at.% in laser mode.

Notably, in nearly all analyses (except for one voltage-pulsed run and two laser-pulsed runs), a peak at 6 Da, attributed to $^{12}C^{2+}$, is observed in the mass-to-charge-state ratio spectra

Atom maps (**Figure 6a**) revealed that the ions of the 1 Da and 2 Da peaks were preferentially located near the $[0002]$ and $[10\bar{1}2]$ crystallographic poles of the HCP lattice of α-Ti in voltage runs, while laser runs did not display a similarly pronounced pole preference. Statistical analysis across both pulsing modes indicated a homogeneous element distribution, with no evidence of clustering or segregation. For all datasets, both CSR values and their corresponding field intensities displayed uniform spatial distribution. The average CSR value was calculated as $(6 \pm 5) \times 10^{-4}$ in voltage runs, corresponding to a field of 34.00 ± 0.62 V/nm, while in laser runs it was (6 ± 9) $\times 10^{-4}$ with a field of 33.92 ± 0.16 V/nm.

### 3.4. Thermal desorption analysis of tritium charged samples

TDA was performed following tritium charging to verify the globally retained tritium content in the whole sample. Immediately following tritium charging, the sample mass was measured to be 0.1268 g. Using the sample mass, the detector counts, and the background signal from the subsequent TDA measurement, a background-corrected specific activity curve was generated along with its integral (representing the total amount of disintegration). These curves, together with the applied temperature ramp, are shown in **Figure 7** as a function of the analysis time. Over the 11 hours of this experiment, the total number of disintegrations counted was $1.96 \times 10^5$ which corresponds to $2.1 \times 10^9$ tritium atoms. We emphasize that this value should be interpreted qualitatively, as detailed in Section 2.4.

The release curve nevertheless provides clear evidence of tritium uptake and meaningful insight into the release behavior. As illustrated in **Figure 7**, the background-corrected specific activity signal was detected at baseline levels until the temperature reached approximately 500 °C (black dashed line 1). Beyond this temperature, a small but noticeable increase in the signal was observed, and after an additional ~30 min, corresponding to a temperature of roughly 700 °C (black dashed line 2), a marked change in slope indicated a pronounced increase in the tritium release rate.

### 3.5. Elemental constituents of the sample after tritium charging based on APT measurements

APT data for the tritium-charged samples were first analyzed excluding the 1, 2, and 3 Da peaks. Across all post-charging intervals, the titanium content, minor impurities, and $^1$H- and $^2$H-related ions remained similar to those in the as-prepared material, with their concentrations reported in **Table 2.** While the oxygen content measured 1 day after charging is similar to that of the as-prepared condition, slightly higher average oxygen contents are observed after 7 and 150 days of storage. However, these differences remain within experimental uncertainty and do not indicate a statistically significant increase. The oxygen and TiO signals were uniformly distributed across all tips, with the exception of a single tip measured in the 1-day post-charging condition (see **Supplementary Figure S3**).

In contrast to the as-prepared samples, a distinct peak at 3 Da was detected in all post-charging states (**Figure 5b,c,d**), except for one tip in the 150-day post-charging interval. The ion-weighted averages of these peaks, together with the total hydrogen content, are summarized in **Table 2**.

Spatial distribution analysis of the hydrogen related ions was similar to that of the as-prepared laser runs, showing no pronounced pole preference. Notably, atoms corresponding to the 3 Da peak were located within the bulk volume of the tip rather than at the surface, and this behavior was consistent across all post-charging intervals (**Figure 6c**).

The ion-weighted average CSR were estimated as $(5 \pm 1) \times 10^{-4}$, $(5 \pm 2) \times 10^{-4}$, and $(10 \pm 1) \times 10^{-4}$ for 1 day, 7 days, and 150 days post-charging, respectively. These values correspond to average electric fields of 33.96 ± 0.11 V/nm, 33.85 ± 0.97 V/nm, and 34.62 ± 0.10 V/nm, respectively.

# 4. Discussion

## 4.1. Hydrogen isotope detection using APT

### 4.1.1. On the 1 Da peak

Prior to tritium charging, both APT and ToF-SIMS consistently exhibit a pronounced 1 Da peak. As noted earlier, this signal corresponds to protons ($^{1}H^{+}$) and is widely recognized as mainly arising from environmental contamination and preparation-induced hydrogen uptake(Chen et al., 2020, 2023; Gault et al., 2024; Kolli, 2017b).

In ToF-SIMS depth profiles, $^{1}H^{+}$ appears as a near-surface maximum that gradually decreases with sputtering depth. APT analysis of the as-prepared samples reveals a consistent 1 Da signal, uniformly distributed along the entire tip in both pulsing modes. This apparent spatial difference could be attributed to the fundamentally different probing length scales of ToF-SIMS and APT, as well as to the presence of titanium's native oxide layer at the surface, which is readily detected by ToF-SIMS but most likely removed during APT tip preparation and just slightly build during fast sample transfer. The $^{1}H^{+}$ content was calculated at approximately 2.08 ± 1.37 at.% under voltage pulsing and 3.56 ± 1.37 at.% under laser pulsing, despite the average electric fields being nearly identical for both acquisition modes.

APT measurements performed in laser-pulsed mode of samples after tritium charging verify a $^{1}H^{+}$ content (1 Da and 2 Da) being similar to the state of the as-prepared sample (**Figure 8a**). A slight decrease in the 1 Da signal was observed 1 day after charging, but this falls within experimental uncertainty and thus, is not statistically significant. Nevertheless, it is relevant to assess whether this apparent decrease could arise from variations in the evaporation field (which depends on multiple parameters such as residual gas pressure, tip geometry, temperature, and local electrostatic conditions) or whether it instead reflects intrinsic tip-to-tip variability associated with the local nature of APT measurements. Because the average field was only 0.04 ± 0.19 V/nm lower than the field measured before charging, and the field distribution was uniform both before and after charging, this observed slight decrease in the 1 Da signal cannot be attributed to changes in evaporation conditions. Considering also the relatively limited number of measurements, these observations suggest that the slight reduction in the 1 Da signal after 1 day of charging is most likely due to statistical scatter arising from to tip-to-tip variability.

#### 4.1.2. On the 2 Da peak

ToF-SIMS measurements on the as-prepared specimens resolved the 2 Da region into two peaks, one at ~2.015 Da, consistent with deuterium, and a smaller one at ~2.0168 Da, corresponding to $^1H_2^+$. Both peaks have intensities roughly three orders of magnitude lower than the 1 Da peak.

In APT experiments, the 2 Da peak originates primarily from molecular $^1H_2^+$ ions formed through the direct ionization of adsorbed $^1H_2$, with only a minor contribution from $^2H^+$ owing to its low natural abundance. Under laser-pulsing conditions, the reduced effective electrostatic field decreases the probability that these molecular ions will dissociate into protons (Breen et al., 2020; Chen et al., 2020, 2019; Kim and Seidman, 2015; Tsong et al., 1983). Indeed, prior to tritium charging, the ion-weighted average 2 Da content was estimated at 0.18 ± 0.18 at.% in voltage mode and 0.62 ± 0.12 at.% in laser mode, an increase of more than a factor of three.

Based on the total number of hydrogen atoms detected by APT and the natural isotopic abundance of deuterium, the expected number of deuterium atoms was calculated and compared with the number of atoms contributing to the 2 Da peak (since no detectable peaks were present at 3 or 4 Da). Prior to tritium charging deuterium accounted for approximately 0.20% of the 2 Da peak in voltage-pulsed datasets and approximately 0.10% in laser-pulsed datasets. After tritium charging, a similar deuterium fraction of approximately 0.075% of the 2 Da peak was measured, consistent with the stability of the total 2 Da ion content observed across all conditions (**Figure 8a**). This stability is expected, as parameters influencing residual $^{1}H_2^{+}$ formation (such as laser pulse energy and pulse frequency) were held constant throughout the laser-pulsed APT experiments.

4.2.3. On the 3 Da peak

A signal observed at 3 Da may originate from any single species or a mixture of $^{3}H^{+}$, $(^{1}H\ ^{2}H)^{+}$, $^{3}He^{+}$, and, in the context of APT, $H_3^{+}$. However, prior to tritium charging the likelihood of contributions from $^{3}H^{+}$ and $^{3}He^{+}$ is minimal given their extremely low natural abundances, and any involvement of $(^{1}H^{2}H)^{+}$ is further diminished by the negligible deuterium concentration present in the samples. ToF-SIMS analysis of pristine Ti samples did not detect any peak at 3 Da, indicating that these species were not formed under the measurement conditions.

In APT, $^{1}H_3^{+}$ formation can occur due the intense fields at the specimen apex, though this process is highly sensitive to experimental parameters such as local field strength, laser pulse energy, and tip temperature (Kolli, 2017b). Notably, before tritium charging, neither voltage- nor laser-pulsed APT revealed a 3 Da signal. In particular, the absence of a 3 Da peak in the as-prepared samples under laser-pulsing provides a direct control for the measurements performed after tritium charging and argues against any detectable contribution from $^{1}H_3^+$ under these conditions. The voltage-pulsed measurements provide only complementary evidence, since molecular ion formation is less readily observed under voltage pulsing. Thus, the absence of a 3 Da signal even under laser-pulsing conditions indicates that any contribution from residual hydrogen is below the detection limit. The same experimental conditions were therefore applied to all tritium charged samples, supporting the exclusion of a detectable $^{1}H_3^+$ contribution to the measured 3 Da signal.

Following tritium charging, a distinct 3 Da peak was consistently observed in all post-charging states during laser-pulsed APT experiments, with similar contents across conditions (**Figure 8b**). Tritium incorporation is discussed in **Supplementary Material S4** and confirmed both by the predicted uptake from Sieverts' law (see **Supplementary Material S5**) and by TDA. With contributions from $^{1}H_3^+$ and $(^{1}H^{2}H)^+$ excluded, and $^{3}H^+$ established, the possibility of a $^{3}He^+$ contribution must be considered, since tritium decay produces $^{3}He$.

Tritium ($^{3}H$) undergoes $\beta^-$ decay with a half-life of 12.32 years, transforming into $^{3}He$ according to the nuclear reaction:

$$^{3}H \rightarrow\ ^{3}He + \beta^- + \bar{\nu}_e \qquad (5)$$

By the time of the charging campaigns, approximately 43 % of the initial $^{3}H$ present in the gas mixture during filling of the bottle had decayed, reducing the $^{3}H$ concentration and generating $^{3}He$. The $^{3}He$ content for each campaign was estimated from the difference between the initial $^3H$ concentration (500 ppm) and the remaining $^3H$ and is reported in **Table 3**.

Despite the comparable content fractions of helium and tritium in the gas phase, $^3H$ is expected to dominate the species absorbed into titanium, not only due to its slightly higher concentration but primarily because of its significantly stronger chemical reactivity. In contrast, $^3He$ is chemically inert and, although it may become physically trapped at defect sites at the surface, it does not appreciably dissolve in or migrate through the titanium lattice under the high-temperature (500 °C) charging conditions. The contribution of $^3He$ to the observed 3 Da signal can be therefore considered negligible, and the peak can be attributed to $^3H$. Complementary ToF-SIMS measurements on tritium-charged specimens could provide additional confirmation of the 3 Da assignment, though such experiments were not feasible in the present study due to radiological constraints of the instrument.

With the origin of the 3 Da peak established, the relative tritium abundance was quantified by normalizing the tritium signal to the total detected hydrogen signal (that is 1 Da plus twice the 2 Da peak, since 2 Da mainly corresponds to $H_2^+$ and the deuterium contribution is minimal). The average values were 671, 9794, and 2369 appm for the 1-, 7-, and 150-day post-charging states, respectively. The evolution of the measured tritium-to-hydrogen ratio may be interpreted in terms of the different mobilities of the two isotopes. Owing to the isotope effect, hydrogen diffuses and outgasses more rapidly than tritium, resulting in a preferential loss of hydrogen during the initial period following charging and, consequently, an increase in the measured tritium-to-hydrogen ratio. As the readily mobile hydrogen becomes depleted, its outgassing rate decreases, whereas tritium continues to desorb from the specimen. This leads to a gradual reduction in the tritium-to-hydrogen ratio at longer storage times. Although this interpretation is consistent with the observed temporal evolution, additional experiments would be required to establish the underlying mechanisms and to assess the possible influence of surface oxide layers or trapping sites.

### 4.2. Influence of the oxide layer

Commercially pure titanium (Ti) possesses a natural titanium oxide surface layer typically ranging from 3 to 10 nm in thickness (Li et al., 2024), acting as efficient barrier to hydrogen isotope transport. ToF-SIMS analysis (see **Sec. 3.2**) of the as-prepared samples before tritium charging confirmed the presence of this native near-surface oxide. This oxide layer has been found to play a dual role in our tritium analysis, as illustrated in the following.

The mechanism of hydrogen migration in rutile $TiO_2$ (110) has been modeled and found to involve high activation energies for proton hopping. Namely, approximately 0.60 eV for movement parallel to the c-axis, and about 1.23 eV for motion perpendicular to it (the latter involving both bond rotation and hopping) (Bates et al., 1979). For tritium, the activation energy increases slightly to ~0.69 eV along the c-axis (Bates et al., 1979). These values highlight the kinetic difficulty of hydrogen isotope transport through $TiO_2$ at ambient temperatures. To overcome the barrier imposed by the oxide layer, a charging temperature of 500 °C (773 K) was selected as it enables the reduction of the oxide layer and formation of water due to the presence of hydrogen (Beste and Bufford, 2021). At this elevated temperature, the originally impermeable oxide layer is dissolved and the Ti surface is reported in the literature to be free of oxides (Beste and Bufford, 2021; Efron et al., 1989; Shugard et al., 2012). After tritium charging, the samples were stored in a nitrogen atmosphere (99.2% $N_2$). Lu et al. (Lu et al., 2009) measured approximately 10 ppm $O_2$ in commercially supplied $N_2$ (99.999% $N_2$) and demonstrated that, under flowing $N_2$ (200 sccm) in a gas-tight furnace, Ti can uptake oxygen and develop a near-surface oxide layer during annealing and cooldown. Consequently, re-oxidation during storage and transport of the Ti sample cannot be entirely excluded in the present study.

Consistent with this interpretation, the TDA result of tritium-charged samples shows a delayed onset of tritium release until temperatures of approximately 500 °C, followed by a pronounced tritium release at higher temperatures (see **Figure 7**). This behavior is consistent with the presence of a surface barrier to tritium outgassing, attributed to an oxide layer formed after charging (see **Supplementary Material S5**). In agreement with this, APT measurements show slightly higher average oxygen contents after 7 and 150 days of storage compared to the as-prepared and 1-day conditions, although these differences remain within experimental uncertainty.

APT measurements indicate a slightly lower tritium content in the 1-day post-charging condition. However, considering the experimental uncertainty, the measured contents remain within statistical margins and show no significant variation across the intervals investigated. But the apparent variation may reflect differences in surface condition, as subtle re-oxidation differences during storage can be expected, and sample transport that could modify the near-surface barrier to isotope transport. However, this effect cannot be quantified, here. Additionally, the scatter may arise from small variations between the three independent charging campaigns despite identical nominal procedures. Furthermore, hydrogen isotope diffusion in α-Ti is crystallographically anisotropic, with diffusion coefficients differing by approximately a factor of two between directions parallel and perpendicular to the c-axis (see

**Supplementary Material S4**). Because no crystallographic control was imposed during APT specimen lift-out, individual tips likely sampled grains with different diffusion pathways, contributing as well to tip-to-tip variability in retained tritium.

Future work would involve a combined EBSD – APT – ToF-SIMS workflow to determine grain-specific oxide-layer variability and its impact on hydrogen-isotope retention, complemented by ellipsometry calibration to convert SIMS fluence into a physical depth scale.

## 5. Conclusion

Using a combined approach of atom probe tomography (APT), time-of-flight secondary ion mass spectrometry (ToF-SIMS), and thermal desorption analysis (TDA) on tritium-free and tritium-charged titanium samples, this study demonstrates that tritium is a highly effective isotopic marker for probing hydrogen behavior at the nanoscale. TDA verified the successful incorporation of tritium into the material. APT measurements revealed a clearly resolved mass peak at 3 Da attributable to tritium uptake, which was absent in the as-prepared titanium sample. In contrast to deuterium, whose detection is intrinsically limited by the persistent 2 Da background arising from residual $^{1}H_2^{+}$, tritium yields a distinct and stable 3 Da signal. This signal is not present in uncharged material and is largely insensitive to contributions from chamber-derived hydrogen.

These characteristics make tritium a practical and highly resolvable tracer for identifying hydrogen at very low concentrations in APT analyses, down to the sub-281 appm T/H levels demonstrated in this work. At the same time, we acknowledge that tritium, deuterium, and protium differ in diffusivity, trapping strength, and phase behavior, and that these isotopic differences cannot be fully resolved within the scope of this study. Our use of tritium is therefore motivated by its superior detectability in APT rather than an assumption of complete physical equivalence among hydrogen isotopes.

Overall, this work establishes tritium as a robust and experimentally advantageous probe for mechanistic investigations of hydrogen uptake, retention, and release, enabling reliable nanoscale studies of hydrogen-related phenomena, including hydrogen embrittlement.

## Tables

*Table 1: Ion-weighted average contents of Ti, O, C, Si, and hydrogen-related molecular ions ($H_{MI}$) (in at.%) for the as-prepared and tritium-charged titanium samples obtained by APT. Peaks at 1, 2, and 3 Da were excluded from the calculation. The reported uncertainties correspond to two ion-weighted standard errors.*

| State | | Ti (at.%) | O (at.%) | C (at.%) | Si (at.%) | $H_{MI}$ (at.%) |
|---|---|---|---|---|---|---|
| As prepared | Voltage | 98.4 ± 0.1 | 0.53 ± 0.04 | 0.07 ± 0.03 | 0.084± 0.005 | 0.24 ± 0.04 |
| | Laser | 98.6 ± 0.2 | 0.57 ± 0.08 | 0.36 ± 0.01 | 0.012 ± 0.003 | 0.034 ± 0.06 |
| Post-charging | 1 day | 98.3 ± 0.8 | 0.66 ± 0.85 | 0.14 ± 0.26 | 0.008 ± 0.007 | 0.11 ± 0.20 |
| | 7 days | 97.5 ± 1.2 | 1.84 ± 1.26 | 0.07 ± 0.06 | 0.010 ± 0.004 | 0.25 ± 0.04 |
| | 150 days | 97.7 ± 1.7 | 1.40 ± 1.7 | 0.07 ± 0.02 | 0.018 ± 0.007 | 0.32 ± 0.09 |

*Table 2: Ion-weighted average compositions (at.%) of the total hydrogen content as well as of the 1 Da, 2 Da and 3 Da peaks in the as-prepared and tritium-charged titanium samples. The total hydrogen content includes contributions from the 1 Da,*

*2 Da, and 3 Da peaks as well as hydrogen bearing molecular ions reported in Table1. The reported uncertainty corresponds to two ion weighted standard errors.*

| **State** | | **Total H (at.%)** | **1 Da (at.%)** | **2 Da (at.%)** | **3 Da (at.%)** |
|---|---|---|---|---|---|
| As prepared | Voltage | 2.38 ± 1.5 | 2.08 ± 1.37 | 0.18 ± 0.18 | Not detected |
| | Laser | 4.49 ± 1.6 | 3.56 ± 1.37 | 0.62 ± 0.12 | Not detected |
| Post-charging | 1 day | 3.08 ± 1.13 | 2.34 ± 0.70 | 0.63 ± 0.45 | 0.004 ± 0.011 |
| | 7 days | 4.79 ± 1.30 | 3.74 ± 1.18 | 0.83 ± 0.17 | 0.028 ± 0.022 |
| | 150 days | 4.31 ± 2.58 | 3.45 ± 1.68 | 0.56 ± 0.26 | 0.007 ± 0.006 |

Table 3: Nominal $^3$H and $^3$He content in the gas mixture for each charging campaign.

| **Campaign date** | May 2024 | January 2025 | March 2025 |
|---|---|---|---|
| **$^3$H content** | 281 appm | 260 appm | 254 appm |
| **$^3$He content** | 219 appm | 240 appm | 246 appm |

## Figure titles, legends, and captions

*Figure 1: Mass-to-charge-state ratio spectra of a pure titanium sample obtained by APT experiments in laser mode: (a) before (blue) and (b) 1 day (red), (c) 7 days (green), and (d) 150 days (cyan) after gas charging with a <281 ppm $^3H$ in $^1H$ gas mixture.*

*Figure 2: Inverse pole figure (IPF) orientation map with color key of a Ti sample prior to tritium charging. High angle grain boundaries are indicated in black (misorientation angle ≥ 5°), while low angle grain boundaries are shown in gray (2° ≤ misorientation angle < 5°).*

*Figure 3: (a) Depth profiles of $Cu^-$ (orange), $C^-$ (blue), $H^-$ (red), $TiH^-$ (pink), $^{18}O^-$ (bright green), $TiO^-$ (green) and $^{49}TiOH^-$ (dark green) measured on the titanium sample prior to tritium charging. All signals are normalized to their respective maximum intensities and plotted as a function of sputtering fluence. (b) Depth-integrated ion maps of $C^-$, $16O^-$, $H^-$, $TiH^-$, $TiO^-$, and total ions acquired from the profiled region. Note that $^{48}TiO^-$ was excluded from the $TiO^-$ isotope sum in both the depth profiles and the depth-integrated images due to signal saturation. H denotes $^1H$ (protium) for consistency with the notation used in the main text.*

*Figure 3: (a) ToF-SIMS mass spectra of a Ti sample prior to tritium charging, covering the 0.8–3.2 Da range, obtained from two depth-profiled regions and showing hydrogen-related peaks attributed to environmental contamination. Enlarged view of one representative profile highlighting (b) the 0.97–1.06 Da region, including the $^1H^+$ peak and a signal associated with an electronic artifact, and (c) the ~2.0138–2.0176 Da region, containing contributions from $^1H_2^+$ and $^2H^+$.*

*Figure 4: Mass-to-charge-state ratio spectra (15.5 to 26 Da region) of a pure titanium sample obtained by APT experiments in laser mode: (a) before (blue) and (b) 1 day (red), (c) 7 days (green), and (d) 150 days (cyan) after gas charging with a <281 appm $^{3}H$ in $^{1}H$ gas mixture.*

*Figure 5: Side and top views of atom maps showing Ti atoms, ions at the 1 Da peak, atoms at the 3 Da peak, and the corresponding field distribution. Data are presented for (a) an as-prepared voltage run, (b) an as-prepared laser run, and (c) a laser run performed one day after tritium charging. All atoms except tritium are represented as points, while tritium atoms are depicted as larger red spheres with a radius of 1.25 nm for improved visualization. Tritium atoms were detected only after charging. The laser-incident direction is reported.*

*Figure 6: Background-corrected specific activity (green), its integral (blue), and the applied temperature ramp (red), as a function of analysis time. The dashed lines 1 and 2 correspond to the temperatures at 500 °C and 700 °C, respectively.*

*Figure 7: APT measurement results from the laser-pulsed experiments illustrating the (a) total hydrogen (grey), 1 Da ions (blue), 2 Da ions (green), and (b) 3 Da ions (red) contents (at.%) before and after tritium charging for 1, 7, and 150 days.*

## Acknowledgments

Financial support by the European Research Council (ERC Consolidator Grant, project TRITIME, grant No. 101043969) is acknowledged. Views and opinions expressed are, however, those of the authors only and do not necessarily reflect those of the European Union or the European Research Council. Neither the European Union nor the granting authority can be held responsible for them.

This work was partly carried out with the support of the Karlsruhe Nano Micro Facility (KNMFi, www.knmf.kit.edu), an Open Access Research Infrastructure within the Karlsruhe High Technology Hub at the Karlsruhe Institute of Technology (KIT – The University in the Helmholtz Association, www.kit.edu).

The authors also acknowledge Dr. Lorenzo Rigutti and Dr. Aissatou Diagne, both from Univ. Rouen Normandie, CNRS INSA Rouen Normandie, Normandie Univ, GPM UMR 6634, 76000 Rouen, France, for fruitful and insightful discussions.

## Declaration of interests

The authors declare that they have no known competing financial interests or personal relationships that could have appeared to influence the work reported in this paper.

## Data statement:

*The data underlying this article are available in* Zenodo, at https://dx.doi.org/10.5281/zenodo.19064457.

## Author contributions

- **Maria Vrellou**: Conceptualization, Methodology, Software, Formal analysis, Investigation, Writing - Original Draft, Visualization
- **Alexander Welle**: Review & Editing, Investigation
- **Stefan Wagner**: Review & Editing, Investigation
- **Marco Weber**: Investigation, Methodology
- **Rolf Rolli**: Methodology, Review & Editing, Validation
- **Hans-Christian Schneider**: Review & Editing, Validation
- **Astrid Pundt**: Review & Editing, Validation
- **Xufei Fang**: Methodology, Formal analysis, Investigation, Writing - Review & Editing, Validation, Supervision
- **Christoph Kirchlechner**: Conceptualization, Methodology, Formal analysis, Writing - Review & Editing, Funding acquisition, Validation, Supervision

## Preprint availability